\documentclass[apj,numberedappendix]{emulateapj}
\usepackage{graphics,epsf}
\usepackage{amsmath}                
\usepackage{amsfonts}               
\usepackage{amssymb}                
\usepackage{epsfig}
 \usepackage{epstopdf}
\usepackage{multirow}
\usepackage{graphicx}
\usepackage{float}
\usepackage{color}
\usepackage[para,online,flushleft]{threeparttable}
\usepackage{etoolbox}
\makeatletter
\def \s{~\rm{s}}

\def \K{~\rm{K}}

\def \yr{~\rm{yr}}

\begin{document}

\title{Possible white dwarf progenitors of type Ia supernovae}

\author{Ealeal Bear\altaffilmark{1} and Noam Soker\altaffilmark{1,2}}

\altaffiltext{1}{Department of Physics, Technion -- Israel Institute of Technology, Haifa
32000, Israel; ealealbh@gmail.com; soker@physics.technion.ac.il}
\altaffiltext{2}{Guangdong Technion Israel Institute of Technology, Shantou, Guangdong Province, China}

\begin{abstract}
We examine catalogs of white dwarfs (WDs) and find that there are sufficient number of massive WDs, $M_{\rm WD} \ga 1.35 M_\odot$, that might potentially explode as type Ia supernovae (SNe Ia) in the frame of the core degenerate scenario. In the core degenerate scenario, a WD merges with the carbon-oxygen core of a giant star, and they form a massive WD that might explode with a time delay of years to billions of years. If the core degenerate scenario accounts for all SNe Ia, then we calculate that about 0.2 per cent of the present WDs in the Galaxy are massive.
Furthermore, we find from the catalogs that the fraction of massive WDs relative to all WDs is about 1-3 per cent, with large uncertainties. Namely, five to ten times the required number. If there are many SNe Ia that result from lower mass WDs, $M_{\rm WD} \la 1.3 M_\odot$, for which another scenario is responsible for, and the core degenerate scenario accounts only for the SNe Ia that explode as massive WDs, then the ratio of observed massive WDs to required is even larger. Despite the several open difficulties of the core degenerate scenario, it is our view that this finding leaves the core degenerate scenario as a possible SN Ia scenario, and possibly even a promising SN Ia scenario. 
\end{abstract}


\section{INTRODUCTION}
\label{sec:intro}

There is no consensus on the binary stellar evolutionary rout that might bring a white dwarf (WD) to undergo thermonuclear explosion that completely destroys the WD in type Ia supernovae (SNe Ia). A detailed comparison of the different evolutionary routes requires to classify them into five basic scenarios (see table 1 in \citealt{Soker2018Rev}).
We list them (by alphabetical order) as follows (for very recent reviews that discuss the five scenarios and references to many earlier papers and reviews see \citealt{LivioMazzali2018, Soker2018Rev, Wang2018}).

(1) In the \textit{core-degenerate (CD) scenario} a CO WD merges with the core of a massive asymptotic giant branch (AGB) star at the final stages of the common envelope evolution.
In the last several years the CD scenario has been developed as a separate SN Ia scenario (e.g., \citealt{Kashi2011, Ilkov2013, AznarSiguanetal2015}) that, according to some of these papers, might account for most SNe Ia.
\cite{Wangetal2017}, on the other hand, conducted a population synthesis study and concluded that this scenario can not result in more than about $20\%$ of the total SNe Ia. Note that \cite{Livio2003} considered the core-WD merger to be a rare event rather than a separate main SN Ia scenario.
(2) In the \textit{double degenerate (DD) scenario} two WDs merge (e.g., \citealt{Webbink1984, Iben1984}, most likely in a violent process (e.g., \citealt{Pakmoretal2010, Pakmoretal2011, Liuetal2016}) a long time after the common envelope evolution has ended. The time delay from merger to explosion (merger explosion delay, or MED) allowed and required by this scenario is an open question (e.g., \citealt{LorenAguilar2009, vanKerkwijk2010, Pakmor2013, Levanonetal2015}).
(3) In the \textit{double-detonation (DDet) mechanism} a detonation of a helium-rich layer that was accreted from a companion ignites the CO WD (e.g., \citealt{Woosley1994, Livne1995, Shenetal2018}).
(4) In the \textit{single degenerate (SD) scenario}. A WD accretes a mass from a non-degenerate companion, reaches a mass close to the Chandrasekhar mass limit ($M_{\rm Ch}$), and explodes (e.g., \citealt{Whelan1973, Han2004, Wangetal2009}).  The WD might explode as soon as it reaches close to $M_{\rm Ch}$, or explode later after it loses some of its angular momentum (e.g., \citealt{Piersantietal2003, DiStefanoetal2011, Justham2011}).
(5) The \textit{WD-WD collision (WWC) scenario} involves the collision of two WDs at about their free fall velocity into each other (e.g., \citealt{Raskinetal2009, Rosswogetal2009, Kushniretal2013, AznarSiguanetal2014}).

As studies of SNe Ia are generally biased towards one or the other of these scenarios, in many cases papers do not consider all five scenarios, hence might reach questionable conclusions. One example are SNe Ia that interact with circumstellar matter (CSM), so called SNe Ia-CSM. Some papers do not consider the CD scenario for these SNe Ia-CSM, and hence conclude that they results from the SD scenario. This is questionable, as at least for the SN Ia PTF11kx the CD scenario seems to do better than the SD scenario (e.g., \citealt{Sokeretal2013}).
We view it mandatory to mention and consider all five scenarios, and so we listed them above.

As for the explosion mechanism and nuclear yields, a useful classification is to two groups, WDs that explode with masses near the Chandrasekhar mass limit, $M_{\rm Ch}$ explosions`, and WDs that explode with masses below that mass, `sub-$M_{\rm Ch}$ explosions` (e.g., \citealt{Maguireetal2018}). Generally, the CD scenario and the SD scenarios belong to $M_{\rm Ch}$ explosions, and the DD, DDet, and WWC scenarios belong to the sub-$M_{\rm Ch}$ explosions.

It is our view that the most promising $M_{\rm Ch}$ scenario is the CD scenario, and the most promising sub-$M_{\rm Ch}$ scenario is the DD scenario.
The other three scenarios, according to this view, might account for some peculiar SNe Ia (see Table 1 in \citealt{Soker2018Rev} for the estimated fraction of SNe Ia from each channel).
However, this view is far from being in the consensus. Several other alternatives exist, such as the more complex process of the SD scenario where a CO WD accreted helium from a companion (e.g., \citealt{Wangetal2009}). Under that view the SD scenario is the main $M_{\rm Ch}$ scenario.

In a recent paper \cite{Ashalletal2018} studied two SNe Ia, and concluded that most SNe Ia, including sub-luminous SNe Ia, are consistent with $M_{\rm Ch}$ delayed-detonation explosions. Since the SD scenario encounters several difficulties, some of them severe \citep{Soker2018Rev}, the results of \cite{Ashalletal2018} strengthen the CD scenario.
Although the CD scenario also suffers from several difficulties, such as the open questions on the merger process, on the delay from merger to explosion, and on the ignition process, in the list of difficulties in Table 1 of \cite{Soker2018Rev}, non of them is listed as a severe problem.
\cite{Dhawanetal2018} find stable $^{58}$Ni in the well studied SN 2014J. This and the analysis of \cite{Diamondetal2018} suggest that SN 2014J is also a product of an $M_{\rm Ch}$ scenario, probably the CD scenario \citep{Soker2015}.

In the CD scenario the merger of the CO WD with the core during the common envelope evolution forms a CO WD with a mass close to $M_{\rm Ch}$. Then there is a delay from the merger to the explosion (MED) that can last up to billions of years (e.g., \citealt{Ilkov2012, Ilkov2013}). It now seems that any scenario that accounts for all SNe Ia needs to have some MED \citep{Soker2018Rev}, although not as long as billions of years.

In the present study, we examine whether there is a potential to find these WDs with mass close to $M_{\rm Ch}$ that are `waiting' to explode in the CD scenario.
 There are earlier claims (e.g., \citealt{Briggsetal2018} and references therein) that single WDs that have very strong magnetic fields are the product of the merger of a WD with the core of a giant during the common envelope phase. But the number of WDs with very strong magnetic fields are not sufficient to account for most SNe Ia, and indeed \cite{Briggsetal2018} did not connect them to SNe Ia.

Several studies have looked for binary WDs that might later experience merger, and some then explode in the frame of the DD scenario (e.g., \citealt{BadenesMaoz2012, MaozHallakoun2017, Maozetal2018}). These studies show that the DD scenario might account mainly for sub-$M_{\rm Ch}$ explosions.
This motivates us to search for the progenitors of $M_{\rm Ch}$-explosion SNe Ia in the frame of the CD scenario, e.g., single WDs with masses of $M_{\rm WD} \ga 1.35M_\odot$. We note that a recent study of Gaia observations suggests that there are many WDs that are the result of a merger \citep{Kilicetal2018}.

In section \ref{sec:expected} we estimate the fraction of WDs that we expect to have this high mass. In section \ref{sec:Stat} we review catalogs of WDs to examine whether there is a potential to find such progenitors.
We summarize in section \ref{sec:summary}.

\section{The expected fraction of massive white dwarfs}
\label{sec:expected}

Following recent studies \citep{Heringeretal2017, FriedmannMaoz2018} we take the delay time distribution (DTD) to have a slope of $-1.5$ \citep{Heringeretal2017} to $-1.3$ \citep{FriedmannMaoz2018}. Namely, the explosion rate of SNe Ia relative to the total number of WDs that are formed after a star formation episode is
\begin{equation}
\frac {d N}{dt} = \frac{N_0}{t_0} \left( \frac{t}{t_0} \right)^{-\alpha},
\label{eq:dotN}
\end{equation}
with $1.3 \la \alpha \la 1.5$.
We can estimate the fraction of WDs that we expect to have high mass using two different approaches as follows.

\subsection{First approach}
\label{subsec:1}
Integrating over time gives the fraction of WDs that exploded till time $t$, $N_e(t)$.
Integration to infinity gives the fraction of WDs that have exploded as
$N_e(\infty)=N_0/(\alpha-1)$.
The ratio of the number of progenitors that survive after time $t$ to the total number of WDs that were formed in the star formation episode is given by
\begin{equation}
N_{\rm s} (t) = \frac {N_0}{\alpha-1} \left( \frac{t}{t_0} \right)^{1-\alpha}.
\label{eq:Nsurvived}
\end{equation}
For a constant star formation rate in the Galaxy over the Galactic history, the ratio of the progenitors that survived at present, $t_{\rm now}$, to the number of WDs is given by
\begin{equation}
N_{\rm s,G}=\frac{1}{t_{\rm now}-t_0} \int_{t_0}^{t_{\rm now}}
\frac {N_0}{\alpha-1}
\left( \frac{t_{\rm now}-t^\prime}{t_0} \right)^{1-\alpha}
d t^\prime.
\label{eq:NsGalaxy1}
\end{equation}
Taking $t_{\rm now} = 10^{10} \yr$ and $t_0=10^8 \yr$, and using $t_{\rm now} \gg t_0$, the fraction of WDs progenitors of SNe Ia from all WDs at present is
\begin{equation}
N_{\rm s,G}=\frac {N_0}{\alpha-1} \frac{1}{2-\alpha}
\left( \frac{t_{\rm now}-t_0}{t_0} \right)^{1-\alpha} \simeq 0.0026,
\label{eq:NsGalaxy2}
\end{equation}
where in the second equality we substituted $\alpha=1.4$ for the DTD slope, and $N_0/(\alpha-1)=0.01$ for the total fraction of WDs that explode as SNe Ia over time (larger than the age of the Galaxy).

\subsection{Second approach}
\label{subsec:2}

In the second approach we take the explosion rate of SNe Ia during the age of the Galaxy $t_{\rm now} =10^{10} \yr$. The average time from star formation to explosion is given by
\begin{equation}
\tau=\frac {\int t N_{\rm s} dt}{\int N_{\rm s} dt}
= \frac{2-\alpha}{3-\alpha} t_{\rm now} \simeq 0.4 \times 10^{10} \yr,
\label{eq:tauNew}
\end{equation}
where in the first equality we have substituted equation (\ref{eq:Nsurvived}), and in the second one, we took $\alpha=1.4$ and $t_{\rm now} = 10^{10} \yr$.

The typical explosion rate of SN Ia in the galaxy is therefore $dn/dt \simeq n/\tau$, where $n$ is the number of SN Ia progenitors.
The observed specific rate of SN Ia in our Galaxy is about $\dot n_{\rm obs} \simeq 5.4 \times 10^{-3} \yr^{-1}$ (\citealt{Lietal2011}; also \citealt{Maozetal2014}). Therefore, our expected number of progenitors in the milky way is $n \simeq \tau \dot n_{\rm obs} \simeq 2 \times 10^7$.
The total number of WDs in the Galaxy is estimated as $n_{\rm WD} \approx 1.15 \times 10^{10}$ (table 2 in \citealt{Napiwotzki2009}), from which we derive the expected ratio of progenitors to total number of WD to be $N_{\rm s,G} \simeq n/n_{\rm WD} \approx 0.0018$.
This is similar to the value we obtain in equation (\ref{eq:NsGalaxy2}).

The number can be even lower. \cite{TsebrenkoSoker2015a} estimate $\approx 20 \%$ of SNe Ia explode inside planetary nebulae (termed SNIP for SN inside planetary nebula). If true, then the number of SNe Ia that `participate' in the long time delay is only about $80 \%$ of SN Ia.
This brings the values we derive in the two approaches to be $N{\rm s,G} \approx 0.002$ and $N{\rm s,G} \approx 0.0015$, respectively.
We note that there are large uncertainties. For example, if the number of SNe Ia inside planetary nebulae is lower, then the CD scenario might account for only a fraction of all SNe Ia (e.g., \citealt{Wangetal2017}).

We conclude that within the CD scenario, about $0.2 \%$ of WDs in the Galaxy should be with a mass of $M_{\rm WD} \ga 1.35 M_\odot$.

Our aim in section \ref{sec:Stat} is to find whether existing catalogs of WDs have the  same fraction of massive WDs as we find theoretically. Namely, that the fraction of WDs with masses of $M_{\rm ED} \ga 1.35 M_\odot$ is larger than about $0.2\%$ of all WDs.

\section{Statistics from Catalogs}
\label{sec:Stat}

\begin{figure}
 \centering
 \hskip -1.10 cm
 \includegraphics[trim= 0.0cm 2.0cm 0.0cm 1.5cm,clip=true,width=0.55\textwidth]{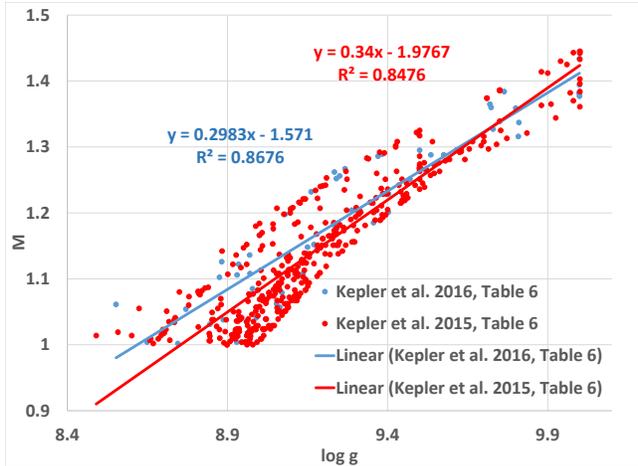}\\
 \vskip 0.1 cm
 \caption{Estimated WD mass ($M_\odot$) vs. $\log(g/{\rm cm} \s^{-2})$ from two catalogs for $ M_{\rm WD} \geq 1M_\odot$. We present the catalog of \cite{Kepleretal2015} with red dots along with a red linear trend-line, and we present the catalog of \cite{Kepleretal2016} with blue dots along with a blue linear trend-line. }
 \label{Mass_Logg}
 \end{figure}

First we find a relation between the log of the WD gravity, $\log g$, and its mass based on two catalogs of WDs from recent years \citep{Kepleretal2015, Kepleretal2016}. We present the mass versus $\log g$ from these two catalogs in Fig. \ref{Mass_Logg}. Although duplicates are present between the catalogs we wanted to show the entire data set in order to deduce a trend-line which we will use later.
We note that the point of SDSS 155758.90+445636.67 with $\log g =8.72$ and $M=7.886M_\odot$, which was out of range, was removed from the sample of \cite{Kepleretal2015}, since clearly the estimated mass is irrelevant.

The \cite{Kepleretal2016} trend-line (blue) and \cite{Kepleretal2015} trend-line (red) (table 6 in both papers) are similar and give $M_{\rm WD} \simeq 1.35$ for $\log g = 9.8$, as we can see directly from the data points.
Hence, we empirically assume that values of $\log g \ga 9.8$ result in almost Chandrasekhar mass, when we consider the Montreal WD catalog.

We review the Montreal WD catalog \citep{Dufouretal2017} which has 30,768 WDs (as of April 2018). We classify the sample by $\log g $ and by the effective temperature. High effective temperatures can indicate mass accretion and hence either the WDs is in a binary system or it is a young WD. Therefore, cool WDs are more relevant to the CD scenario. In table \ref{table:Montreal}, we list the number of WDs in the different classes, including a separation to single WDs and WDs in binary systems.
We make a note of WDs that were found to be in binary systems, marked `binary', and single WDs, marked `SU', in the table.  We note that the binary class are known WDs which are in a binary system while the single/unknown (SU) category can not be separated to real single and those that are not determined to be single. Therefore, this class might include WDs in binaries as well.
The binary category includes binary systems which are composed of MS and a WD and a few systems which are two WDs.
\begin{table}[h!]
\centering
\begin{tabular}{|l|l|l|l|l|l|l|l|}
\hline
$\log g$ & All & $\ge 9.5$ & $\ge 9.6$ & $\ge 9.7$ & $\ge 9.8$ & $\ge 9.9$ & $\ge 10$ \\ \hline
All & 30768 & 541 & 467 & 406 & 368 & 316 & 196 \\ \hline
$10^4 \K$ & 9281 & 306 & 291 & 283 & 274 & 256 & 173 \\ \hline
SU & 28669 & 476 & 410 & 349 & 311 & 262 & 150 \\ \hline
$10^4 \K$;SU & 8750 & 250 & 237 & 229 & 220 & 203 & 127 \\ \hline
Binary & 2099 & 65 & 57 & 57 & 57 & 54 & 46 \\ \hline
$10^4 \K$;B& 531 & 56 & 54 & 54 & 54 & 53 & 46 \\ \hline
\end{tabular}
\caption{Data from the Montreal WD Catalog, by \cite{Dufouretal2017}, arranged by $\log g$ for the following classes. All: all WDs in the catalog. $\le 10 \K$: only WDs with effective temperatures of $T_{\rm eff} < 10^4 \K$. SU: Single/Unknown, meaning single WDs or unknown for being single.
$10^4 \K$;SU: WDs from the SU class with $T_{\rm eff} < 10^4 \K$. Binary: WDs in known binary systems. $10^4 \K$;B: WDs in known binary systems with $T_{\rm eff} < 10^4 \K$.}
\label{table:Montreal}
\end{table}
 The fraction of WDs with $\log g \ga 9.8$, corresponding to $M_{\rm WD} \ga 1.35 M_\odot$, according to table \ref{table:Montreal} is $\simeq 1.2\%$. This ratio is larger, $\simeq 2.9 \%$, when considering only cool WDs, i.e., $T_{\rm eff} \la 10000\K$.
When using only the class of single/unknown WDs the ratio stays the same at about $1\%$ and for binary systems it is $2.7\%$.

\cite{Kepleretal2016} find that 94 WDs in their sample have masses higher than $1M_\odot$, and they suggest that many of them were formed by a merger of two WDs. From figure 3 of \cite{Kepleretal2016} we find that about $4\%$ of all WDs have a mass of $M_{\rm WD} \ga 1.35 M_\odot$ and about $1 \%$ have masses of
$M_{\rm WD} > 1.375 M_\odot$.

 Overall, the fraction of massive WDs, $M_{\rm WD} \ga 1.35 M_\odot$, that we find in these catalogs and the study of \cite{Kepleretal2016}, is about $1-3 \%$. This fraction is about five to ten times the number that is required according to the CD scenario for SNe Ia, about $0.2 \%$ (section \ref{sec:Stat}). The fraction of WDs with masses of $M_{\rm WD} \ga 1.38 M_\odot$ is about $1\%$, namely, about five times the fraction required by the CD scenario.

Our finding of about $5-10$ times the required number of massive WDs in the CD scenario is encouraging, because several other channels can form massive WDs. Some, but not many, massive WDs might be formed by a single star evolution that leaves a massive ONe WD. Many massive WDs might be formed by the merger of an ONe WD with another WD, during the common envelope evolution or later by emitting gravitational waves (e.g., \citealt{Canalsetal2018, {Kashyapetal2018}}). Another channel is a mass transfer from a companion, a process that might form both CO massive WDs and ONe-rich massive WDs, in the frame of the single degenerate scenario but with a long time delay to explosion (e.g., \citealt{MengHan2018}).

\section{SUMMARY}
\label{sec:summary}

As the community is far from any consensus on the evolutionary routs that brings WDs to explode as SNe Ia (section \ref{sec:intro}), any study that can shed some light on one or more of the five scenarios is welcome. In the present study we have looked into the question of whether there are enough potential SN Ia progenitors in the frame of the CD scenario.

If the CD scenario accounts for all SNe Ia, then about $0.2 \%$ of all WDs in the Galaxy at present should have a mass that will bring the WD to explode, e.g., very close to the limiting Chandrasekhar mass (section \ref{sec:expected}). We found from the catalogs of WDs (section \ref{sec:Stat}) that the fraction of WDs with masses of $M_{\rm WD} \ga 1.35 M_\odot$ is about $1-3\%$, while the fraction of
WDs with masses of $M_{\rm WD} \ga 1.38 M_\odot$ is about $1\%$.
One must take into account that there are large uncertainties in deriving the value of $\log g$, and hence there are large uncertainties in the derived fraction of WDs with the required mass.

Nonetheless, we find that the potential number of massive enough WDs is about $5-10$ times the fraction of massive WDs required by the CD scenario. Many of these are WDs rich with Ne, namely ONe WDs. In most cases these will not lead to regular SNe Ia, but rather, most of these WDs end in a collapse to a neutron star, with a minority that might lead to peculiar SNe Ia.
 We do note that there are suggestions that hybrid CONe WDs do actually lead to regular SNe Ia (e.g., \citealt{Denissenkovetal2013, Wangetal2014}), hence further study to better determine the masses of these WDs and their composition is required of course.

If two scenarios or more account for SNe Ia, then the constraint on the number of massive WDs becomes weaker. For example, if the DD scenario accounts for the long time delay from star formation to explosion and for sub-$M_{\rm Ch}$ explosions, while the CD scenario accounts for $M_{\rm Ch}$ explosions with shorter time delay, say with about half of SNe Ia coming from each of these two scenarios, then the required fraction of massive WDs progenitors of SNe Ia becomes less than $0.1 \%$. In that case the estimated fraction of massive WDs from observations is more than ten times the required fraction.

We can summarize by stating that despite the several problems and disadvantages of the CD scenario (that include the open questions of, e.g., the long delay from merger to explosion and the ignition process; see \citealt{Soker2018Rev}),  our main finding leaves the CD scenario as a viable and promising SN Ia scenario.

We thank an anonymous referee for detailed and very helpful comments.
This research was supported by the Asher Fund for Space Research at the Technion, and the Israel Science Foundation.


\begin{thebibliography}

\bibitem[Ashall et al.(2018)]{Ashalletal2018} Ashall, C., Mazzali, P.~A., Stritzinger, M.~D., et al.\ 2018, \mnras, 477, 153

\bibitem[Aznar-Sigu{\'a}n et al.(2015)]{AznarSiguanetal2015} Aznar-Sigu{\'a}n, G., Garc{\'{\i}}a-Berro, E., Lor{\'e}n-Aguilar, P., Soker, N., \& Kashi, A.\ 2015, \mnras, 450, 2948

\bibitem[Aznar-Sigu{\'a}n et al.(2014)]{AznarSiguanetal2014} Aznar-Sigu{\'a}n, G., Garc{\'{\i}}a-Berro, E., Magnien, M., \& Lor{\'e}n-Aguilar, P.\ 2014, \mnras, 443, 2372

\bibitem[Badenes \& Maoz(2012)]{BadenesMaoz2012} Badenes, C., \& Maoz, D.\ 2012, \apjl, 749, L11


\bibitem[Briggs et al.(2018)]{Briggsetal2018} Briggs, G.~P., Ferrario, L., Tout, C.~A., \& Wickramasinghe, D.~T.\ 2018,  \mnras, 478, 899

\bibitem[Canals et al.(2018)]{Canalsetal2018}  Canals, P., Torres, S., \& Soker, N.\ 2018, arXiv:1806.06730

\bibitem[Denissenkov et al.(2013)]{Denissenkovetal2013} Denissenkov, P.~A., Herwig, F., Truran, J.~W., \& Paxton, B.\ 2013, \apj, 772, 37

\bibitem[Dhawan et al.(2018)]{Dhawanetal2018} Dhawan, S., Fl{\"o}rs, A., Leibundgut, B.,  Maguire, K., Kerzendorf, W., Taubenberger, S., Van Kerkwijk, M.~H., \& Spyromilio, J.\ 2018, arXiv:1805.02420

\bibitem[Diamond et al.(2018)]{Diamondetal2018} Diamond, T.~R., Hoeflich, P., Hsiao, E.~Y., et al.\ 2018, \apj, 861, 119 (arXiv:1805.03556)

\bibitem[Di Stefano et al.(2011)]{DiStefanoetal2011} Di Stefano, R., Voss, R., \& Claeys, J.~S.~W.\ 2011, \apjl, 738, LL1

\bibitem[Dufour et al.(2017)]{Dufouretal2017} Dufour, P., Blouin, S., Coutu, S., et al.\ 2017, 20th European White Dwarf Workshop, 509, 3

\bibitem[Friedmann \& Maoz(2018)]{FriedmannMaoz2018} Friedmann, M., \& Maoz, D.\ 2018, \mnras, 479, 3563 (arXiv:1803.04421)

\bibitem[Han \& Podsiadlowski(2004)]{Han2004} Han, Z., \& Podsiadlowski, P.\ 2004, \mnras, 350, 1301

\bibitem[Heringer et al.(2017)]{Heringeretal2017} Heringer, E., Pritchet, C., Kezwer, J., Graham, M.~L., Sand, D., \& Bildfell, C.\ 2017, \apj, 834, 15

\bibitem[Iben \& Tutukov(1984)]{Iben1984} Iben, I., Jr., \& Tutukov, A.~V.\ 1984, \apjs, 54, 335

\bibitem[Ilkov \& Soker(2012)]{Ilkov2012} Ilkov, M., \& Soker, N.\ 2012, \mnras, 419, 1695

\bibitem[Ilkov \& Soker(2013)]{Ilkov2013} Ilkov, M., \& Soker, N.\ 2013, \mnras, 428, 579

\bibitem[Justham(2011)]{Justham2011}  Justham, S.\ 2011, \apjl, 730, LL34

\bibitem[Kashi \& Soker(2011)]{Kashi2011} Kashi, A., \& Soker, N.\ 2011, \mnras, 417, 1466

\bibitem[Kashyap et al.(2018)]{Kashyapetal2018}  Kashyap, R.,Lor{\'e}n-Aguilar, P., Haque, T., Garc{\'{\i}}a-Berro, E., \& Fisher, R.\ 2018

\bibitem[Kepler et al.(2015)]{Kepleretal2015} Kepler, S.~O., Pelisoli, I., Koester, D., et al.\ 2015, \mnras, 446, 4078

\bibitem[Kepler et al.(2016)]{Kepleretal2016} Kepler, S.~O., Pelisoli, I., Koester, D., et al.\ 2016, \mnras, 455, 3413

\bibitem[Kilic et al.(2018)]{Kilicetal2018} Kilic, M., Hambly, N.~C., Bergeron, P., Genest-Beaulieu, C., \& Rowell, N.\ 2018, \mnras, 479, L113 (arXiv:1805.01227)

\bibitem[Kushnir et al.(2013)]{Kushniretal2013} Kushnir, D., Katz, B., Dong, S., Livne, E., \& Fern{\'a}ndez, R.\ 2013, \apjl, 778, L37

\bibitem[Levanon et al.(2015)]{Levanonetal2015} Levanon, N., Soker, N.,\& Garc{\'{\i}}a-Berro, E.\ 2015, \mnras, 447, 2803

\bibitem[Li et al.(2011)]{Lietal2011} Li, W., Chornock, R., Leaman, J., et al.\ 2011, \mnras, 412, 1473

\bibitem[Liu et al.(2016)]{Liuetal2016} Liu, D.-D., Wang, B., Podsiadlowski, P., \& Han, Z.\ 2016, \mnras, 461, 3653

\bibitem[Livio \& Mazzali(2018)]{LivioMazzali2018} Livio, M., \& Mazzali, P.\ 2018, Physics Reports, 736, 1 (arXiv:1802.03125)

\bibitem[Livio \& Riess(2003)]{Livio2003} Livio, M., \& Riess, A.~G.\ 2003, \apjl, 594, L93

\bibitem[Livne \& Arnett(1995)]{Livne1995} Livne, E., \& Arnett, D.\ 1995, \apj, 452, 62

\bibitem[Lor{\'e}n-Aguilar et al.(2009)]{LorenAguilar2009} Lor{\'e}n-Aguilar, P., Isern, J., \& Garc{\'{\i}}a-Berro, E.\ 2009, \aap, 500, 1193

\bibitem[Maguire et al.(2018)]{Maguireetal2018} Maguire, K., Sim, S.~A., Shingles, L., et al.\ 2018, \mnras, 477, 3567

\bibitem[Maoz \& Hallakoun(2017)]{MaozHallakoun2017} Maoz, D., \& Hallakoun, N.\ 2017, \mnras, 467, 1414

\bibitem[Maoz et al.(2018)]{Maozetal2018} Maoz, D., Hallakoun, N., \& Badenes, C.\ 2018, \mnras, 476, 2584

\bibitem[Maoz et al.(2014)]{Maozetal2014} Maoz, D., Mannucci, F., \& Nelemans, G.\ 2014, \araa, 52, 107

\bibitem[Meng \& Han(2018)]{MengHan2018}  Meng, X.-C., \& Han, Z.-W.\ 2018, \apjl, 855, L18

\bibitem[Napiwotzki(2009)]{Napiwotzki2009} Napiwotzki, R.\ 2009, Journal of Physics Conference Series, 172, 012004

\bibitem[Pakmor et al.(2011)]{Pakmoretal2011} Pakmor, R., Hachinger, S., R{\"o}pke, F.~K., \& Hillebrandt, W.\ 2011, \aap, 528, A117

\bibitem[Pakmor et al.(2013)]{Pakmor2013} Pakmor, R., Kromer, M., \& Taubenberger, S.\ 2013, arXiv:1302.2913


\bibitem[Pakmor et al.(2010)]{Pakmoretal2010} Pakmor, R., Kromer, M., R{\"o}pke, F.~K., Sim, S. A., Ruiter, A. J. and Hillebrandt, W.\ 2010, \nat, 463, 61


\bibitem[Piersanti et al.(2003)]{Piersantietal2003} Piersanti, L.,
Gagliardi, S., Iben, I., Jr., \& Tornamb{\'e}, A.\ 2003, \apj, 583, 885

\bibitem[Raskin et al.(2009)]{Raskinetal2009} Raskin, C., Timmes, F.~X., Scannapieco, E., Diehl, S., \& Fryer, C.\ 2009, \mnras, 399, L156

\bibitem[Rosswog et al.(2009)]{Rosswogetal2009} Rosswog, S., Kasen, D., Guillochon, J., \& Ramirez-Ruiz, E.\ 2009, \apjl, 705, L128

\bibitem[Shen et al.(2018)]{Shenetal2018} Shen, K.~J., Kasen, D., Miles, B.~J., \& Townsley, D.~M.\ 2018, \apj, 854, 52

\bibitem[Soker(2015)]{Soker2015} Soker, N.\ 2015, \mnras, 450, 1333

\bibitem[Soker(2018)]{Soker2018Rev} Soker, N.\ 2018, Science China Physics, Mechanics, and Astronomy, 61, 49502

\bibitem[Soker et al.(2013)]{Sokeretal2013} Soker, N., Kashi, A., Garcia-Berro, E., Torres, S., \& Camacho, J.\ 2013, \mnras, 431, 1541

\bibitem[Tsebrenko \& Soker(2015)]{TsebrenkoSoker2015a} Tsebrenko, D., \& Soker, N.\ 2015, \mnras, 447, 2568

\bibitem[van Kerkwijk et al.(2010)]{vanKerkwijk2010} van Kerkwijk, M.~H., Chang, P., \& Justham, S.\ 2010, \apjl, 722, L157

\bibitem[Wang(2018)]{Wang2018} Wang, B.\ 2018, RAA 2018, 18, 49 (arXiv:1801.04031)


\bibitem[Wang et al.(2009)]{Wangetal2009} Wang, B., Meng, X., Chen, X., \& Han, Z.\ 2009, \mnras, 395, 847

\bibitem[Wang et al.(2014)]{Wangetal2014} Wang, B., Meng, X., Liu, D.-D., Liu, Z.-W., \& Han, Z.\ 2014, \apjl, 794, L28

\bibitem[Wang et al.(2017)]{Wangetal2017} Wang, B., Zhou, W.-H., Zuo, Z.-Y.Li, Y.-B., Luo, X., Zhang, J.-J., Liu, D.-D. and Wu, C.-Y.\ 2017, \mnras, 464, 3965

\bibitem[Webbink(1984)]{Webbink1984} Webbink, R.~F.\ 1984, \apj, 277, 355

\bibitem[Whelan \& Iben(1973)]{Whelan1973} Whelan, J., \& Iben, I., Jr.\ 1973, \apj, 186, 1007

\bibitem[Woosley \& Weaver(1994)]{Woosley1994} Woosley, S.~E., \& Weaver, T.~A.\ 1994, \apj, 423, 371

\end{thebibliography}
\end{document}